\documentstyle[aps,pre]{revtex}
\begin{document}
\draft
\title{Linear response of vibrated granular systems to sudden changes in the
 vibration intensity}
\author{J. Javier Brey and A. Prados}
\address{F\'{\i}sica Te\'{o}rica, Universidad de Sevilla, Apartado de Correos
 1065, 41080, Sevilla, Spain}
\date{today}
\maketitle
\begin{abstract}
The short-term memory effects recently observed in vibration-induced
compaction of granular materials are studied. It is shown that they can be
explained by means of quite plausible hypothesis about the mesoscopic
description of the evolution of the system. The existence of a critical
time separating regimes of ``anomalous'' and ``normal'' responses is
predicted. A simple model fitting into the general framework is analyzed
in the detail. The relationship between this work and previous studies
is discussed.

\end{abstract}
\pacs{PACS numbers: 45.70.Cc, 61.43.Fs, 81.05.Rm}

\section{Introduction}
\label{s1}
Experiments have shown that when a loose packing of grains is submitted to
vertical vibration or ``tapping'' slowly approaches a steady state of higher
packing fraction \cite{KFLJyN95,NKBJyN98}. The final steady density is a
decreasing function of the dimensionless parameter characterizing the
vibration intensity. Moreover, the relaxation is slower for smaller
vibration intensity. In the time evolution of the
system neither convection effects nor oscillatory behaviour are observed.
The study of the kinetics of compaction is important both from a formal point
of view and because of its economical relevance in many industrial
processes. Most of the peculiar behaviours exhibited by granular materials
submitted to  vibration or tapping processes show a great similarity with
conventional structural glasses. This includes slow relaxation, annealing
properties, and hysteresis effects.

The first study of the response of a granular system to a sudden change in the
vibration intensity we are aware of, was carried out by means of numerical
simulations of a model for compaction \cite{Ni99}, and the data indicated the
presence of memory effects in the evolution of the density of the system. Very
recently \cite{JTMyJ00}, memory effects were also directly observed in a series
of experiments. The  results showed that the system has a short-term memory of
its shaking history, so that the response in the evolution of the density to a
change in the vibration intensity at a given time, is not determined by the
density at that time. Mathematically, this phenomenon implies that the time
evolution of the density does not obey a closed ordinary first order
differential equation.

In this paper, we propose a general theoretical framework to understand the
origin and characteristic features of the memory effects seen both in
simulations and in experiments. Using quite plausible hypothesis, we will be
able to explain the short-time response of the system to a small change in the
vibration intensity. In particular, the theory predicts that a decrease
(increase) in the intensity can lead to an increase (decrease) of the compaction
rate on short time scales, in agreement with experiments. Nevertheless, this has
not to be necessarily so. If the change in the intensity is made at the early
stages of the compaction process, the theory we will develop leads to a
modification of the compaction rate having the same sign as the intensity
change. In fact, there is a critical time, which depends on the tapping
intensity before the change, separating the regions of ``normal'' and
``anomalous'' responses. The existence of these two different regimes has not
been verified experimentally up to now, although such a behaviour has been
numerically observed in a simple model for granular compaction \cite{He00}.

As an illustration of the theory, we discuss its application to a model for
compaction introduced recently \cite{BPyS99,BPyS00}. The model has already been
showed to reproduce the qualitative behaviour of granular materials under
tapping. Here we will show that it also captures the same short-term memory
effects seen in the experiments. Moreover, it fits perfectly in the general
scheme developed in this paper, therefore providing a first test of validation
of the ideas in which the theory is based. We have also used the model to
investigate the relaxation of the system following a perturbation in the
vibration intensity for a short time period. This idea also originates from the
experiments reported in ref. \cite{JTMyJ00}. The results indicate that the
response function is accurately described by a Kohlrausch-Williams-Watts (KWW)
or stretched exponential function.

The paper is organized as follows. In the next Section, some general
properties of the equation governing the time evolution of the density in
tapping processes of granular media are discussed. These properties are
used in Sec.\ \ref{s3} to analyze the short-term memory effects by
considering the response of the system to a small change in the vibration
intensity. The theory is particularized for a simple model for tapping
in Sec. \ref{s4}, where other patterns of change of the vibration intensity
are also considered. The choices were originated from the experiments
reported in ref. \cite{JTMyJ00}. Finally, Sec.\ \ref{s5} contains some
additional comments and final remarks, as well as a relevant discussion of the
relationship of our work to previous experimental and theoretical studies.

\section{Evolution of the density in discrete tapping processes}
\label{s2}

Let $\Gamma$ denote the dimensionless parameter characterizing the intensity
of the vibration applied to the granular medium. In typical experiments
\cite{KFLJyN95}, $\Gamma$ is defined as the ratio of peak acceleration of a
tap to the gravity $g$. Under very general conditions, the time evolution of
the density $\rho$ in a discrete tapping process will be given at a mesoscopic
level by an equation of the form
\begin{equation}
\label{2.1}
\dot{\rho}\equiv \frac{d \rho(t)}{dt}=f_{1}(\Gamma) \mu_{1}(t)-f_{2}(\Gamma)
\mu_{2}(t).
\end{equation}
Here the time is measured in units of complete taps in some continuous limit,
$f_{1}(\Gamma)$ and $f_{2}(\Gamma)$ are semidefinite positive functions of
$\Gamma$ having dimensions of frequency, and $\mu_{1}(t)$ and $\mu_{2}(t)$
are positive quantities depending on the state of the system, but they
are not univocally determined by the density at the same instant $\rho(t)$.
Therefore, Eq.\ (\ref{2.1}) is not in general a closed equation and can not
be solved by itself. The two terms on the right hand side of the equation
describe elementary processes increasing and decreasing the density,
respectively.

The structure of Eq.\ (\ref{2.1}) as a gain-loss equation is consistent with the
experimental observations in compaction processes, as we will discuss in detail
in the following. Also, if the elementary events taking place in the system
being vibrated can be described by means of a master equation, a formal equation
like this follows directly. This is the case for some simple kinetic models for
compaction introduced recently \cite{KyB94,BKNJyN98,BPyS99,BPyS00}.

Since Eq.\ (\ref{2.1}) describes the evolution of the density as a
consequence of tapping, the functions $f_{1}$ and $f_{2}$ must vanish in the
limit of no tapping $\Gamma=0$, so that
\begin{equation}
\label{2.2}
f_{1}(0)=f_{2}(0)=0.
\end{equation}
Because of continuity, it follows, at least for small values of the intensity
$\Gamma$, that
\begin{equation}
\label{2.3}
f^{\prime}_{1}(\Gamma) \equiv \frac{d}{d \Gamma} f_{1}(\Gamma) >0, \quad
f^{\prime}_{2}(\Gamma) \equiv \frac{d}{d \Gamma} f_{2}(\Gamma) >0.
\end{equation}
We will assume that the above inequalities hold for arbitrary $\Gamma$. The
physical reason for this assumption is that we expect the number of elementary
processes taking place in the system to increase as $\Gamma$ increases. Of
course, this does not imply by itself that the rate of variation of the density
also increases. The behaviour of $\dot{\rho}$ depends on the net balance between
the gain and loss elementary events, as indicated by Eq.\ (\ref{2.1}). This
picture is in agreement with the qualitative role of temperature played by the
shaking intensity in many different aspects
\cite{NKBJyN98,BPyS00,PByS00,EyO89,EyG98,BKLyS00}.

In the long time limit of a tapping process with constant $\Gamma$, the
experiments show that the system reaches a steady state with a density
$\rho_{s}$, which is a monotonic decreasing function of $\Gamma$, as displayed
by the ``reversible'' branch in cycling experiments \cite{BKNJyN98}. Let us
point out that the relaxation process is very slow, and for very small values of
$\Gamma$ the steady density is hard to reach within the experimental time scale.
Therefore, the function $\rho_{s}(\Gamma)$ verifies that
$d\rho_{s}(\Gamma)/d\Gamma <0$, and it is bounded by the two formal limits
\begin{equation}
\label{2.4}
\rho_{min}=\lim_{\Gamma \rightarrow \infty} \rho_{s}(\Gamma), \quad
\rho_{max}=\lim_{\Gamma \rightarrow 0} \rho_{s}(\Gamma).
\end{equation}
Particularization of Eq.\ (\ref{2.1}) for a steady state yields
\begin{equation}
\label{2.5}
f_{1}(\Gamma) \mu_{1s}=f_{2}(\Gamma)\mu_{2s},
\end{equation}
where $\mu_{1s}$ and $\mu_{2s}$ denote the steady values of the quantities
$\mu_{1}$ and $\mu_{2}$, respectively. As pointed out above, $\mu_{1}(t)$ and
$\mu_{2}(t)$ are not expected to be simply functions of $\rho(t)$ in general.
But, on the other hand, it seems sensible to assume that the steady state
reached by a given system in a tapping experiment is fully determined by the
intensity $\Gamma$ or, equivalently, by $\rho_{s}$. Therefore, we assume that
$\mu_{1s}$ and $\mu_{2s}$ are functions of $\rho_{s}$, and in the following we
are going to investigate some qualitative properties of these functions. For
$\rho_s \rightarrow \rho_{min}$, $\mu_{2s}(\rho_s)$ must vanish, since by
definition at $\rho_s=\rho_{min}$ there are no processes decreasing the density.
Therefore, it is
\begin{equation}
\label{2.6}
\lim_{\rho_s \rightarrow \rho_{min}} \mu_{2s}(\rho_s)=0, \quad
\lim_{\rho_s \rightarrow \rho_{min}} \mu_{1s}(\rho_s) >0.
\end{equation}
The second relationship expresses that starting from the density $\rho_{min}$,
any tapping process of arbitrary intensity $\Gamma$ can only produce an
increase of the density. What happens in the steady high density limit?. A
similar argument to the one carried out above would lead to
\begin{equation}
\label{2.7}
\lim_{\rho_s \rightarrow \rho_{max}} \mu_{1s}(\rho_s)=0, \quad
\lim_{\rho_s \rightarrow \rho_{max}} \mu_{2s}(\rho_s)>0.
\end{equation}
Nevertheless, some care is required when analyzing this limit. Simple models for
discrete tapping lead to an absorbent steady state in the high density limit
\cite{BPyS99,BPyS00}. That means that the system will not be able to leave this
state when submitted to tapping of arbitrary intensity. This is equivalent to
say that $\mu_{2s}(\rho_s)$ also vanishes for $\rho_s \rightarrow \rho_{max}$.
As a consequence, and in order to include such a possibility in our formulation,
instead of Eq.\ (\ref{2.7}) we will assume the more general and precise
condition
\begin{equation}
\label{2.8}
\lim_{\rho_s \rightarrow \rho_{\max}} \frac{\mu_{1s}(\rho_s)}{\mu_{2s}(\rho_s)}=0
\, ,
\end{equation}
i.\ e., $\mu_{2s}>>\mu_{1s}$ when $\rho_s\rightarrow\rho_{\max}$, and the
density loss term is dominant in that limit. Let us note that Eqs.\ (\ref{2.6})
yield
\begin{equation}
\label{2.9}
\lim_{\rho_s \rightarrow \rho_{\min}} \frac{\mu_{1s}(\rho_s)}{\mu_{2s}(\rho_s)}=
\infty.
\end{equation}

The simplest behaviour that is consistent with Eqs.\ (\ref{2.8}) and (\ref{2.9})
is that the ratio $\mu_{1s}/\mu_{2s}$ be a monotonic decreasing function of the
steady density $\rho_s$ going from infinity to zero. Since there is not any
physical reason to expect a more complicated density dependence, we assume this
is the case in our formalism. From the steady condition given by Eq.\
(\ref{2.5}), it follows that
\begin{equation}
\label{2.10}
\frac{\mu_{1s}}{\mu_{2s}}=\frac{f_{2}(\Gamma)}{f_{1}(\Gamma)} \equiv
 g(\Gamma).
\end{equation}
The function $g(\Gamma)$ is a measure of the rate of the decompaction processes
with respect to the rate of the compaction ones. Because of Eqs.\ (\ref{2.8})
and (\ref{2.9}), it is
\begin{equation}
\label{2.11}
\lim_{\Gamma \rightarrow \infty}g(\Gamma) =\infty, \quad
\lim_{\Gamma \rightarrow 0}g(\Gamma)=0.
\end{equation}
Taking derivative with respect to the intensity $\Gamma$ in Eq.\ (\ref{2.10})
it is obtained:
\begin{equation}
\label{2.12}
\frac{dg(\Gamma)}{d\Gamma}=\frac{d\rho_{s}}{d \Gamma} \frac{d}{d \rho_{s}}
\left( \frac{\mu_{1s}}{\mu_{2s}} \right)  >0,
\end{equation}
where we have taken into account the monotonically decreasing  density
dependence of $\mu_{1s} /\mu_{2s}$ assumed above. The physical meaning of
Eq.\ (\ref{2.12}) is evident: the rate of the decompaction processes
grows faster with $\Gamma$ than the rate of the compaction processes. Also
this implication of our assumptions seems physically plausible.

In summary, we can write the equation for the time evolution of the
density in discrete tapping processes as
\begin{equation}
\label{2.13}
\frac{d \rho(t)}{dt}=f_{1}(\Gamma) \left[ \mu_{1}(t)-g(\Gamma) \mu_{2}(t)
\right],
\end{equation}
with $f_{1}$ and $g$ being positive increasing functions of $\Gamma$, both of
them vanishing in the limit $\Gamma \rightarrow 0$. The quantities $\mu_{1}(t)$
and $\mu_{2}(t)$ are some moments of the complete distribution function of the
system, and they contain the influence of correlations on the evolution of the
density. As a consequence, (\ref{2.13}) is not a closed equation.

Because $g(\Gamma)$ vanishes for $\Gamma \rightarrow 0$, if a tapping experiment
with low enough intensity $\Gamma$ is carried out, the decompaction term
$f_{1}(\Gamma) g(\Gamma) \mu_2(t)$ will be negligible in the first stages of the
process, i.\ e.,
\begin{equation}
\frac{\mu_1(t)}{g(\Gamma)\mu_2(t)} \gg 1        \, ,
\end{equation}
and the evolution of the system will be approximately described by
\begin{equation}
\label{2.14}
\frac{ d \rho(t)}{dt} \simeq f_{1}(\Gamma) \mu_{1}(t).
\end{equation}
At much later times, when $\rho(t)$ is close enough to the asymptotic steady
value, the decompaction contribution in Eq.\ (\ref{2.13}) plays a decisive role,
leading to a steady density $\rho_s < \rho_{\max}$, and it is
\begin{equation}
\label{2.15}
\frac{\mu_{1}(t)}{g(\Gamma)\mu_{2}(t)} = O(1).
\end{equation}
The observed behaviour that the system tends towards a steady state and,
therefore, a regime where $\mu_{1}$, $\mu_{2}$, and $g$ verify a relationship of
the form given in Eq.\ (\ref{2.15}), can be understood if $\mu_{1}(t)$ decreases
in time while $\mu_{2}$ increases. Quite interestingly, this is consistent with
a mean field approximation in which $\mu_{1}(t)$ is replaced by
$\mu_{1s}[\rho(t)]$ and $\mu_{2}(t)$ by $\mu_{2s}[\rho(t)]$. Since $\rho(t)$
increases monotonically in time, and $\mu_{1s}/\mu_{2s}$ is a monotonic
decreasing function of the density, it follows that the left hand side of Eq.\
(\ref{2.15}) will decay in time.

Of course, as long as Eq.\ (\ref{2.14}) is accurate, the larger $\Gamma$ the
faster the compaction of the system, in agreement with experiments. Over a
larger time scale, the complete Eq.\ (\ref{2.14}), including the decompaction
term, is needed in order to explain the dependence of the steady density
$\rho_{s}$ on $\Gamma$, and also the existence of a a slow long time tail in the
relaxation of the density, once $\rho_{s}-\rho(t)$ is very small. In this
context, the presence of an ``anomalous'' density relaxation, following an
inverse logarithm law, would be associated to some specific dynamical properties
of the compaction term $\mu_{1}$ when the system is submitted to ``nonlinear''
tapping processes \cite{KFLJyN95,NKBJyN98}. We use the term ``nonlinear'' here
in the sense that, in the experiments, the initial value of the density is not
very close to the steady density.

Later on, we will show that an evolution equation like Eq.\ (\ref{2.13}) applies
in the case of a simple model recently introduced  to describe discrete tapping
\cite{BPyS99,BPyS00,PByS00}. Another similar equation is obtained for the
``parking'' model \cite{KyB94,BKNJyN98,KNyT99,TTyV00}, although this latter
refers to continuous vibration processes, in which the system is not allowed to
relax to a metastable configuration between every two vibration cycles. In the
parking model, the state of maximum density $\rho_{\max}$ is not totally
absorbent, but this possibility has been included in our theory, as discussed
below Eq.\ (\ref{2.7}). By identifying the intensity of tapping $\Gamma$ with
the ratio between the desorption and adsorption rates in the parking model, it
is trivial to check by using the expressions in Ref.\ \cite{KyB94} that the
quantities corresponding to $\mu_{1s}$ and $\mu_{2s}$ verify that their ratio is
an increasing function of $\rho_{s}$, and also the limiting behaviour given in
Eqs.\ (\ref{2.8}) and (\ref{2.9}). Moreover, the steady density is an increasing
function of the quantity playing the role of the vibration intensity. In
conclusion, the parking model belong to the general class of systems we have
considered.

\section{Response to small vibration intensity jumps}
\label{s3}

In this Section we will investigate whether  Eq.\ (\ref{2.13}), which has been
built under very general arguments and is expected to have a wide range of
applicability, is able to predict the memory effects recently observed in
vibration-induced compaction in granular materials \cite{JTMyJ00}.  The fact
that the equation is not closed for the density, implies that its time
evolution in a given experiment with constant $\Gamma$ is not determined
by its the initial value. Starting from the same value $\rho_{0}$,
different time evolutions are possible depending on the way in which the
system was prepared. Our aim is to analyze some particular relevant
manifestations of this general statement.

Consider that, starting from a given configuration, the system is tapped with an
intensity $\Gamma$. At a certain time $t_{w}$, the intensity is instantaneously
changed to $\Gamma + \Delta \Gamma$. Quite peculiarly, it has been observed in
the experiments that the change in the compaction rate has opposite sign that
$\Delta \Gamma$ on short time scales, though in the long time regime the
relaxation is slower for smaller values of the intensity of vibration $\Gamma$.
The same kind of effect has also been previously found numerically in some
models for compaction \cite{He00,TTyV00a,Ni99}, although it only shows up when
the time interval $t_{w}$ is not too short. If $\Gamma$ is changed at the
beginning of the compaction process, the variation of the compaction rate has
the same sign as $\Delta \Gamma$ \cite{He00,note}.

Application of Eq.\ (\ref{2.13}) for the instant $t_{w}^{-}$, just before
the change in the intensity of the vibration, yields
\begin{equation}
\label{3.1}
r_{w} \equiv \dot{\rho}(t_{w}^{-})=f_{1}(\Gamma) \left[ \mu_{1w}^{-}
-g(\Gamma) \mu_{2w}^{-} \right],
\end{equation}
where $\mu_{1w}^{-}=\mu_{1}(t_{w}^{-})$ and  $\mu_{2w}^{-}=\mu_{2}(t_{w}^{-})$.
When the intensity of vibration is changed into $\Gamma + \Delta \Gamma$, the
compaction rate becomes
\begin{equation}
\label{3.2}
r^{\prime}_w\equiv \dot{\rho} (t_{w}^{+})= f_{1}(\Gamma +\Delta \Gamma)
\left[ \mu_{1w}^{+} -g(\Gamma+\Delta \Gamma) \mu_{2w}^{+} \right].
\end{equation}
The continuity of the distribution function of the system implies that
$\mu_{1}^{-}=\mu_{1}^{+}$ and $\mu_{2}^{-}=\mu_{2}^{+}$ for an instantaneous
jump of $\Gamma$, although there is a discontinuity $\Delta r_{w} =
r_{w}^{\prime}-r_{w}$ in the compaction rate. For $\Delta \Gamma$ small we
can approximate
\begin{equation}
\label{3.3}
\frac{\Delta r_{w}}{\Delta \Gamma} =f^{\prime}_{1} (\Gamma)\left[
\mu_{1w}-g(\Gamma) \mu_{2w} \right] -f_{1}(\Gamma) g^{\prime} (\Gamma)
\mu_{2w} =\frac{f^{\prime}_{1}(\Gamma)}{f_{1}(\Gamma)} r_{w}-f_{1}(\Gamma)
g^{\prime} (\Gamma)\mu_{2w}.
\end{equation}
Therefore, if over the compaction curve corresponding to intensity $\Gamma$
we define the function
\begin{equation}
\label{3.4}
\lambda(t)=
\frac{f^{\prime}_{1}(\Gamma)}{f_{1}(\Gamma)} r(t)-f_{1}(\Gamma)g^{\prime}
(\Gamma)\mu_2(t),
\end{equation}
the sign of this function at the time $t_{w}$ when the intensity is changed will
determine the relative behaviour of $\Delta r_w$ with respect to $\Delta
\Gamma$, for infinitesimal changes of the latter. If $\lambda_{w} \equiv
\lambda (t_{w}) <0$, the anomalous response observed in the experiments will
follow, while if $\lambda_{w} >0$ the compaction rate will change in the
same direction as $\Delta \Gamma$. Let us analyze the sign of the function
$\lambda(t)$. In the long time limit, formally $t_{w} \rightarrow \infty$,
the system is known to reach the asymptotic steady density, so that
$r_{w} \rightarrow 0$ and, consequently,
\begin{equation}
\label{3.5}
\lambda_{\infty} =\lim_{t\rightarrow \infty} \lambda(t)=
-f_{1}(\Gamma)g^{\prime}(\Gamma) \mu_{2s}(\Gamma)<0,
\end{equation}
where it has been taken into account that both $f_{1}(\Gamma)$ and
$g_{1}(\Gamma)$ are positive increasing functions of $\Gamma$ and that
$\mu_{2s}(\Gamma)>0$.

On the other hand, if the initial density in the experiment is the minimum
possible density of the system at rest $\rho_{min}$, corresponding to the random
loose packing configuration, it follows from the properties of $\mu_2$ that
\begin{equation}
\label{3.6}
\lim_{t\rightarrow 0} \lambda (t)=
\frac{f^{\prime}_{1}(\Gamma)}{f_{1}(\Gamma)}\, r(0) >0,
\end{equation}
Even though we have considered in our discussion that $\rho(t=0)=\rho_{min}$
in order to derive the above inequality, the same result will apply if the
initial density is close enough to it, so that the first term on the right
hand side of Eq.\ (\ref{3.4}) dominates the second one in the initial regime.

Then, we conclude that for short times $\Delta r_{w}$ and $\Delta \Gamma$
have the same sign, while for large times their signs are opposite. This
renders compatible and explains what is seen in the experiments and also in
numerical studies of simple models. From our analysis it follows that there
is (at least) a time $t_{c}$, which depends on the value of $\Gamma$, such
that the response of the system to a small variation of the intensity of
tapping is qualitatively different for $t<t_{c}$ and $t>t_{c}$.

The study carried out in this Section has been restricted to small instantaneous
changes in $\Gamma$, allowing the use of a linear analysis of Eq.\ (\ref{2.13}).
Whether the behaviour of the system remains the same when submitted to a finite
change in the shaking intensity, it can not be inferred from our analysis. In
this case, nonlinear effects can modify dramatically the response of the system.
More will be said about this in the next section of the paper.

\section{Application to a simple model for compaction}
\label{s4}
The general scenario developed in the previous sections will be
particularized here for a one-dimensional lattice model for compaction
\cite{BPyS00,PByS00}. In the model, each site $i$ can be either empty
or occupied by a particle. A variable $m_{i}$ is defined, being $m_{i}=1$
in the former case and $m_{i}=0$ in the latter. A configuration of the
system is fully specified by giving the values of all the variables
${\bf m} \equiv \{m_{i} \}$. As usual, we will refer to the empty sites as
being occupied by a hole.

Let us describe the dynamics of the system when submitted to a
discrete tapping process. Mechanical stability requires that all the
holes be isolated, i.e. surrounded by two particles, at the end of
every tap. The time evolution of the system is defined as a Markov process,
and formulated by means of a master equation for the probability distribution
of the system \cite{BPyS00,PByS00}. The equation contains the transition
rates $W({\bf m}|{\bf m}^{\prime})$ from state ${\bf m}^{\prime}$ to state
${\bf m}$. There are three kinds of possible transitions. Indicating only the
variables associated to the sites involved in the transitions, the
nonvanishing transition rates are:
\begin{enumerate}
\item
Elementary diffusive events conserving the number of particles,
\begin{equation}
\label{4.1}
W(010|100)=W(010|001)=\frac{\alpha}{2},
\end{equation}
\item
Transitions increasing the number of particles,
\begin{mathletters}
\label{4.2}
\begin{equation}
\label{4.2a}
W(010|101)=\frac{\alpha}{2},
\end{equation}
\begin{equation}
\label{4.2b}
W(001|101)=W(100|101)=\frac{\alpha}{4}.
\end{equation}
\end{mathletters}
\item
Transitions increasing the number of holes, i.e. decreasing
the number of particles,
\begin{mathletters}
\label{4.3}
\begin{equation}
\label{4.3a}
W(01010|00100)=\frac{\alpha^{2}}{2},
\end{equation}
\begin{equation}
\label{4.3b}
W(01010|01000)=W(01010|00010)=\frac{\alpha^{2}}{4}.
\end{equation}
\end{mathletters}
\end{enumerate}
In the above equations, $\alpha$ is a positive constant, characterizing the
tapping process completely, and playing in the model a role similar to the
intensity of vibration $\Gamma$ in real experiments. For $\alpha \neq 0$, the
system evolves from any arbitrary initial configuration to a final steady
state with density
\begin{equation}
\label{4.4}
\rho_{s}(\alpha)=\frac{1}{2} \left[ 1+ \left( 1+4\alpha \right)^{-1/2}
\right].
\end{equation}
From here it follows that
\begin{equation}
\label{4.5}
\lim_{\alpha \rightarrow 0} \rho_{s}=1\equiv\rho_{max}, \quad
\lim_{\alpha \rightarrow \infty} \rho_s=\frac{1}{2}\equiv \rho_{min},
\end{equation}
being
\begin{equation}
\label{4.6}
\frac{d\rho_{s}}{d \alpha} <0,
\end{equation}
for all $\alpha$. Therefore the density in the model has the same kind of
dependence on the intensity $\alpha$ as assumed in the general discussion in
Sec.\ \ref{s2}. The time evolution of $\rho$ is obtained from the master
equation for the model, and reads \cite{ByPu}
\begin{equation}
\label{4.7}
\dot{\rho}= \alpha x_{101}(t)-\frac{\alpha^{2}}{2} \left[ x_{00100}(t)
+\frac{1}{2}x_{01000}(t)+\frac{1}{2}x_{00010}(t) \right],
\end{equation}
where $x_{010}$ is the concentration of three-site clusters of the form
hole-particle-hole, $x_{00100}$ is the concentration of five-site clusters
formed by a hole between two pairs of particles, and so on. Comparison of
Eqs. (\ref{2.1}) and (\ref{4.7}) allows to identify
\begin{equation}
\label{4.8}
f_{1}(\alpha)=\alpha, \quad f_{2}(\alpha)=\alpha^{2},
\end{equation}
\begin{equation}
\label{4.9}
\mu_{1}(t)=x_{101}(t), \quad
\mu_{2}(t)=\frac{1}{2} x_{00100}(t)+\frac{1}{4} \left[
 x_{01000}(t)+x_{00010}(t) \right].
\end{equation}

In the steady state, the only correlations in the system are those forbidding to
have two nearest-neighbour holes \cite{ByPu}. Then, it is a simple matter to
compute the steady values of the several cluster concentrations appearing in
Eq.\ (\ref{4.9}) with the result
\begin{equation}
\label{4.10}
\mu_{1s}=\frac{(1-\rho_{s})^{2}}{\rho_{s}},
\end{equation}
\begin{equation}
\label{4.11}
\mu_{2s}=\frac{(1-\rho_{s})(2\rho_{s}-1)^{2}}{\rho_{s}^{2}}\, .
\end{equation}
In the limit $\rho_{s} \rightarrow \rho_{min}=1/2$,
\begin{equation}
\label{4.12}
\mu_{1s} \rightarrow \frac{1}{2}, \quad
\mu_{2s} \rightarrow 0,
\end{equation}
while in the high density limit $\rho_{s} \rightarrow   \rho_{max}=1$, both
$\mu_{1s}$ and $\mu_{2s}$ vanish, as a consequence of the absorbent character of
the state with all the sites occupied by particles. The ratio
\begin{equation}
\label{4.13}
\frac{\mu_{1s}(\rho_s)}{\mu_{2s}(\rho_s)}=\frac{\rho_{s}(\rho_{s}-1)}{(2\rho_{s}
-1)^{2}}
\end{equation}
vanishes in this latter limit. Equations (\ref{4.12}) and (\ref{4.13}) are in agreement
with Eqs.\ (\ref{2.6}) and (\ref{2.8}). Moreover,
$\mu_{1s}(\rho_s)/ \mu_{2s}(\rho_s)$ is a monotonic decreasing function of
$\rho_s$ and, consistently (see Eq.\ (\ref{2.10})),
\begin{equation}
\label{4.14}
g(\alpha)=\frac{f_{2}(\alpha)}{f_{1}(\alpha)}=\alpha,
\end{equation}
is an increasing function of $\alpha$, vanishing in the limit $\alpha
\rightarrow 0$.

We conclude that this model for compaction fits perfectly the general picture
developed in the previous Sections. Equation (\ref{2.13}) particularized for
the model is
\begin{equation}
\label{4.15}
\frac{d\rho(t)}{dt}=\alpha (\mu_{1}-\alpha \mu_{2} ),
\end{equation}
with $\mu_{1}$ and $\mu_{2}$ defined in Eq.\ (\ref{4.9}). To solve Eq.\
(\ref{4.15}) we would need some (approximate) expressions for the cluster
concentrations as functions of the density.

If we submit the system to the tapping experiment described in Sec.\ \ref{s3},
the effect of the intensity change $\Delta \alpha$ at $t=t_{w}$ on the compact
rate will be given by
\begin{equation}
\label{4.16}
\frac{\Delta r_{w}}{\Delta \alpha}=\frac{r_{w}}{\alpha}-\alpha \mu_{2w},
\end{equation}
in the limit of small $\Delta \alpha$. Therefore, the function determining
whether the response of the system will be ``normal'' or ``anomalous'' is
\begin{equation}
\label{4.17}
\lambda(t)=\frac{r(t)}{\alpha}-\alpha \mu_{2}(t).
\end{equation}
In Fig.\ \ref{f1} this function is plotted for $\alpha=0.15$. The curve has been
obtained by Monte Carlo simulation of the master equation of the system. The
data represent an average over $10$ different runs. The initial state was the
one corresponding to the steady minimum density. For this particular value of
the intensity $\alpha$, $\lambda(t)$ changes sign between taps $19$ and $20$,
i.e. $ 19\leq t_{c} \leq 20 $. For comparison purposes, we have also plotted the
mean field approximation for $\lambda (t)$ (dashed line). The latter has been
constructed by substituting in Eq.\ (\ref{4.17}) $\mu_1(t)$ and $\mu_2(t)$ by
$\mu_{1s}(\rho(t))$ and $\mu_{2s}(\rho(t))$, respectively, and using for the
density the simulation results. It is seen that the mean field approximation
also changes sign, but for larger times, and it is always above the ``exact''
Monte Carlo curve. This is consistent with the mean field approximation giving a
faster approach to the steady state than the actual relaxation of the system
\cite{ByPu}.

According to the results derived in this paper, $\Delta r_{w}$ is expected to
have different sign that $\Delta  \alpha$ for $t_{w}>t_{c}$ and the same for
$t_{w}<t_{c}$. In order to check this theoretical prediction, we have carried
out series of Monte Carlo simulations, all them starting in the minimum density
configuration, with $\alpha=0.15$. At $t_{w}=50$, the value of the intensity
$\alpha$ was instantaneously changed to $\alpha^{\prime}$. The results for four
different values of $\alpha^{\prime}$ are reported in Fig.\ \ref{f2}, namely
$0.1, 0.125, 0.15, 0.175,$ and $0.2$, from top to bottom. The central value
corresponds to no change. Since in these simulations it is $t_{w}>t_{c}$, the
compaction rate is observed to decrease as the value of $\alpha^{\prime}$
increases. It is also seen that the amplitude of the jump in the compaction rate
is larger for $\alpha^{\prime}=0.2$ ($\Delta \alpha =0.05$) than for
$\alpha^{\prime}=0.1$ ($\Delta \alpha =
-0.05$). This feature can not be explained by Eq.\ (\ref{4.16}), and it is
 due to nonlinear effects that have been neglected in the linear approximation
used here. This will be analyzed below.

In Fig.\ \ref{f3} the same set of experiments is carried out, with the
only difference that in this case the intensity $\alpha$ is modified
at $t_{w}= 10 <t_{c}$. The several curves correspond to the same values as
in Fig.\ \ref{f2}, but now they are ordered from bottom to top. As
predicted by the theory, the variation of the compaction rate has the same
sign as the change in $\alpha$. Moreover, the same kind of nonlinear
effects as in Fig.\ \ref{f2} are present.

Now, we will briefly discuss the nonlinear corrections in $\Delta\alpha$ to the
change in the compaction rate. It is easy to show that
\begin{equation}
\label{4.18}
\Delta r_w=\Delta\alpha \lambda_w -(\Delta\alpha)^2 \mu_{2w} \, .
\end{equation}
The second term   on the rhs of Eq.\ (\ref{4.18}) is neglected in the linear
approximation. In this simple model, the nonlinear correction is always
negative, so it can modify dramatically the response of the system to the jump
$\Delta\alpha$ if the linear term $\Delta\alpha \lambda_w > 0$. In particular,
there is a critical value
\begin{equation}
\label{4.19}
\Delta\alpha_c=\frac{\lambda_w}{\mu_{2w}}
\end{equation}
such that $\Delta r_w=0$. For smaller jumps, $|\Delta\alpha|<|\Delta\alpha_c|$,
the sign of $\Delta r_w$ is the one predicted by the linear approximation, but
for larger jumps, $|\Delta\alpha|>|\Delta\alpha_c|$, the sign of $\Delta r_w$ is
the opposite to the prediction of the linear approximation. For the sake of
concreteness, in Fig.\ \ref{f6} we have repeated the numerical experiment of
Fig.\ \ref{f2}, but with larger intensity jumps. From the Monte Carlo
simulation, we obtain the critical value $\Delta\alpha_c\simeq -0.1$ for
$\alpha=0.15$ and $t_w=50$. Then, at $t_w=50$ we change the vibration intensity
from $\alpha=0.15$ to $\alpha^\prime=\alpha+\Delta\alpha_c=-0.05$, finding that
the compaction rate does not change in the short time limit $t-t_w\rightarrow
0$. Moreover, if the vibration intensity is further decreased,
$\alpha^\prime=0.03$, the compaction rate also decreases, while the linear
approximation predicted an increase of the compaction rate if
$\alpha^\prime<\alpha$, since $\lambda_w<0$.

Following Ref.\ \cite{JTMyJ00}, we have also considered another series of
numerical experiments where the system was tapped up to the same density with
three different intensities, $\alpha=0.2$, $0.15$, and $0.1$, respectively.
Afterwards, the system was always tapped with the same intensity
$\alpha^{\prime}=0.15$. The time evolution of the density is shown in Fig.\
\ref{f4}, where the time origin for each experiment has been taken at the time
when the system reached the prescribed density, namely $\rho=0.8$. The figure
clearly shows that the evolution of the density for $t>0$ strongly depends on
the previous tapping history, indicating the relevance of short-term memory
effects. Mathematically, this is equivalent to say that $\mu_{1}(t)$ and
$\mu_{2}(t)$ in Eq.\ (\ref{2.1}) are not determined univocally by the density at
the same time, so that it is not in fact a closed first order ordinary
differential equation. Note that in all the plotted curves the jump in the
compaction rate has opposite sign than the variation of the intensity. We have
verified that $\lambda(t)$ is negative at the time in which the intensity is
modified in all cases, the behaviour being then consistent with the theory.

\section{Discussion}
\label{s5}
Along this paper, we have studied the non-equilibrium linear response of a
vibrated granular system to an instantaneous change in the intensity of the
taps. In the first part, a general theory was developed on the basis of some
plausible hypothesis about the mesoscopic dynamics of the system. The results
are in qualitative agreement with the experimental observations. In particular,
the presence of short-term memory effects appears as correlated with the
relaxation properties of the system at constant intensity. An important
theoretical prediction, not observed in the experiments yet, is the existence of
a critical time $t_{c}$. For times $t<t_{c}$ the response of the system to a
change in the intensity is ``normal'', in the sense that an increase in the
intensity produces a positive jump in the compaction rate. On the other hand,
for $t>t_{c}$ an ``anomalous'' response is produced. The change in the
compaction rate has opposite sign that the modification of the vibration
intensity, in contrast with the long time behaviour found in experiments, where
the relaxation is faster for larger vibration intensity.

In the second part of the paper, a simple model for compaction has been
considered. It is shown to fit perfectly into the general scheme developed
before, allowing a detailed quantitative analysis of the theoretical predictions.
This is not a peculiarity of this model, since the ``parking'' model
\cite{BKNJyN98} also verifies all the conditions assumed in the theoretical
framework. In fact, this is not surprising because this latter model has a
mathematical structure very similar to the one considered in this paper
\cite{BPyS99}.

In Sec.\ \ref{s4} we have shown that our model reproduces the experimentally
observed behaviour of the system when submitted to changes in the vibration
intensity under different conditions \cite{JTMyJ00}. Now we will refer to a more
complicated pattern of changes in the intensity that are also discussed in ref.
\cite{JTMyJ00}. First, the system is shaken with an intensity $\Gamma_{0}$
($\alpha_{0}$ in the model notation) for a long period of time, so that the
system practically reaches a steady density. Afterwards, at a time taken as
$t=0$, the intensity is switched to $\Gamma > \Gamma_{0}$ for a  given period of
time $t_{0}$ and, finally, the system is tapped again with the original
intensity $\Gamma_{0}$, and the subsequent relaxation of the system is studied.
Experimentally it was found that the relaxation is slower the larger $t_{0}$,
the system ``ages''. Moreover, on the basis of a simple two-state model, it was
proposed that
\begin{equation}
\label{5.1}
\rho(t)-\rho_{s} \sim  \frac{\exp (-\kappa_{0}t)}{t},
\end{equation}
for $t-t_{0} \gg 1$. In the above expression $\kappa_{0}$ is a decreasing
function of $t_{0}$. Josserand {\em et al.} \cite{JTMyJ00} also reported that
the relaxation can be fitted by a superposition of exponentials, all of them
with the same amplitude. We have carried out numerically this kind of
experiments in our model. In Fig.\ \ref{f5} we present the results obtained with
$\alpha_{0}=0.3$, $\alpha=0.5$, and four different values of $t_{0}$, namely
$t_{0}=1,2,4$, and $8$ from bottom to top. The plotted response function
$\phi(t)$ is defined as
\begin{equation}
\label{5.2}
\phi (t)=\frac{\rho_{s}(\alpha_{0})-\rho(t)}{\rho_{s}(\alpha_{0})-
\rho(t_{0})},
\end{equation}
where $\rho(t_{0})$ is the density of the system at the time in which the the
intensity is switched back to $\alpha_{0}$. For the model, the steady values of
the density can be computed analytically  \cite{BPyS99}, and it is
$\rho_{s}(\alpha=0.3)\simeq 0.8371$. For a given time interval $t-t_0$,
$\phi(t)$ increases with the ``waiting'' time $t_0$. Thus, the relaxation is
slower for larger $t_0$, consistently with the experimental observation
\cite{JTMyJ00}. Also, we have fitted (solid lines) the data to a stretched
exponential or KWW function \cite{WW70},
\begin{equation}
\label{5.3}
\rho_{KWW}(t) =\rho_{s}(\alpha_{0}) -\left[\rho_{s}(\alpha_{0})-\rho(t_{0})
\right] \exp \left[ - \left( \frac{t-t_{0}}{\tau} \right)^{\beta} \right],
\end{equation}
with $\tau$ and $\beta$  being fitting parameters. As observed in the figure,
the fit is quite satisfactory, except for times very close to $t_{0}$ and,
probably, for very large times. The parameter $\beta$ in Eq.\ (\ref{5.3})
measures the width of the relaxation time distribution. he values we have found
go from $\beta=0.366$ for $t_{0}=1$ to $\beta=0.478$ for $t_{0}=8$. The latter
is close to the value $1/2$, characteristic of systems whose dynamics is
dominated by one-dimensional diffusive processes. However, the KWW relaxation is
not equivalent to a superposition of exponentials with the same amplitude, as
proposed in \cite{JTMyJ00}. Therefore, this point deserves more work in the
future, both theoretically and experimentally. With respect to the long time
behaviour predicted by Eq.\ (\ref{5.1}) we could not reach a definite answer.
Although the numerical data seem to be compatible with it, the noise is too
large and further high-precision studies  would be required.

Finally, a crucial point in the analysis presented in this paper is the small
amplitude of the perturbation in the vibration intensity $\Delta \Gamma$. As
pointed out at the end of Sec.\  \ref{s3}, the behaviour following a large
change in the intensity may be different. In the model considered in Sec.\
\ref{s4} the nonlinear corrections are very simple, leading always to a decrease
in the compaction rate and to the appearance of a critical value of the
intensity jump, such that no change in the vibration intensity is observed in
the short time regime. Moreover, for jumps larger than the critical one, the
sign of the change in the compaction rate is reversed as compared with the
prediction of the linear approximation. We think that it is worth looking for
this kind of behaviour in other models for compaction, and also in experiments
with real granular systems.

\acknowledgments
This research has been partially supported by the Direcci\'{o}n General de
Investigaci\'{o}n Cient\'{\i}fica y T\'{e}cnica (Spain) through Grant No.
PB98-1124.

\begin{figure}
\caption{
Time evolution of the function $\lambda$ defined in Eq.\ \protect\ref{4.17}, for
a vibration intensity $\alpha=0.15$ (solid line). Also plotted is the mean field
approximation for $\lambda$ (dashed line).}
\label{f1}
\end{figure}

\begin{figure}
\caption{
Evolution of the density, $\rho$, as a function of the number of taps, t. Five
numerical experiments are shown, the vibration intensity $\alpha$ was changed at
$t_w=50$,  where $\lambda<0$, from $0.15$ to $0.1$, $0.125$, $0.15$, $0.175$,
and $0.2$, from top to bottom. Thus, the central curve corresponds to no change
in the tapping intensity $\alpha$ (solid line), while the dotted and dashed
lines correspond to a decrease and an increase in $\alpha$, respectively. The
``anomalous'' response experimentally observed shows up.}
\label{f2}
\end{figure}

\begin{figure}
\caption{
The same experiment of Fig.\ \protect\ref{f2}, but the change in intensity is
introduced at an earlier time $t_w=10$, at  which $\lambda>0$. The curves
correspond to the same values of $\alpha^\prime$ as in Fig.\ \protect\ref{f2},
but now are ordered from bottom to top. In this region the response is
``normal'', i.\ e., the compaction rate increases with the vibration intensity.}
\label{f3}
\end{figure}

\begin{figure}
\caption{
Time evolution of the density $\rho$ when the vibration intensity $\alpha$ is
changed at $t_w=50$, where $\lambda<0$, from $\alpha$=0.15 to
$\alpha^\prime=0.05$ (circles) and $0.03$ (squares), respectively. The curve
corresponding to a constant vibration intensity $\alpha=0.15$ is plotted for
reference (solid line). For such large jumps, the linear approximation is not
valid, and the compaction rate do not increase. In fact, $\alpha^\prime=0.05$
corresponds to the critical value $\Delta\alpha_c$, for which no change in the
compaction rate is observed for short times.}
\label{f6}
\end{figure}

\begin{figure}
\caption{
Time evolution of the density for a system which was tapped up to the same
density $\rho=0.8$ using three different intensities, $\alpha=0.2$ (circles),
$0.15$ (squares), and $0.1$ (triangles). Afterwards, the system was always
tapped with $\alpha^{\prime}=0.15$.  The time origin for each experiment has
been taken at the time when the system reached the prescribed density, namely
$\rho=0.8$. The evolution for $t>0$ strongly depends on the pre-history of the
system.}
\label{f4}
\end{figure}

\begin{figure}
\caption{
Relaxation function $\phi(t)$,  defined in Eq.\ (\protect\ref{5.2}), of the
model, when it is prepared by tapping for a long time with $\alpha=0.3$, and
afterwards tapped for $t_0=1$, $2$, $4$, and $8$ (from bottom to top) with a
larger intensity $\alpha^\prime=0.5$. Finally, the intensity is turned back to
the original intensity $\alpha=0.3$. All the curves tend to zero in the infinite
time limit, and the solid lines are the best numerical fits to a KWW function. }
\label{f5}
\end{figure}


\begin{references}
\bibitem{note}
In the lower graph in Fig.\  2 of ref.\ \ \protect{\cite{Ni99}}, it is observed
that for $t-t_{w} \rightarrow 0$, when the intensity of the vibration is
increased the height of the system decreases, i.\ e., the density increases.
This graph corresponds to the lowest intensity considered, so that the waiting
time $t_{w}$ is much smaller than the characteristic relaxation time. This
explains the change in behaviour as compared with the other two plots in the
same figure.
\end{references}
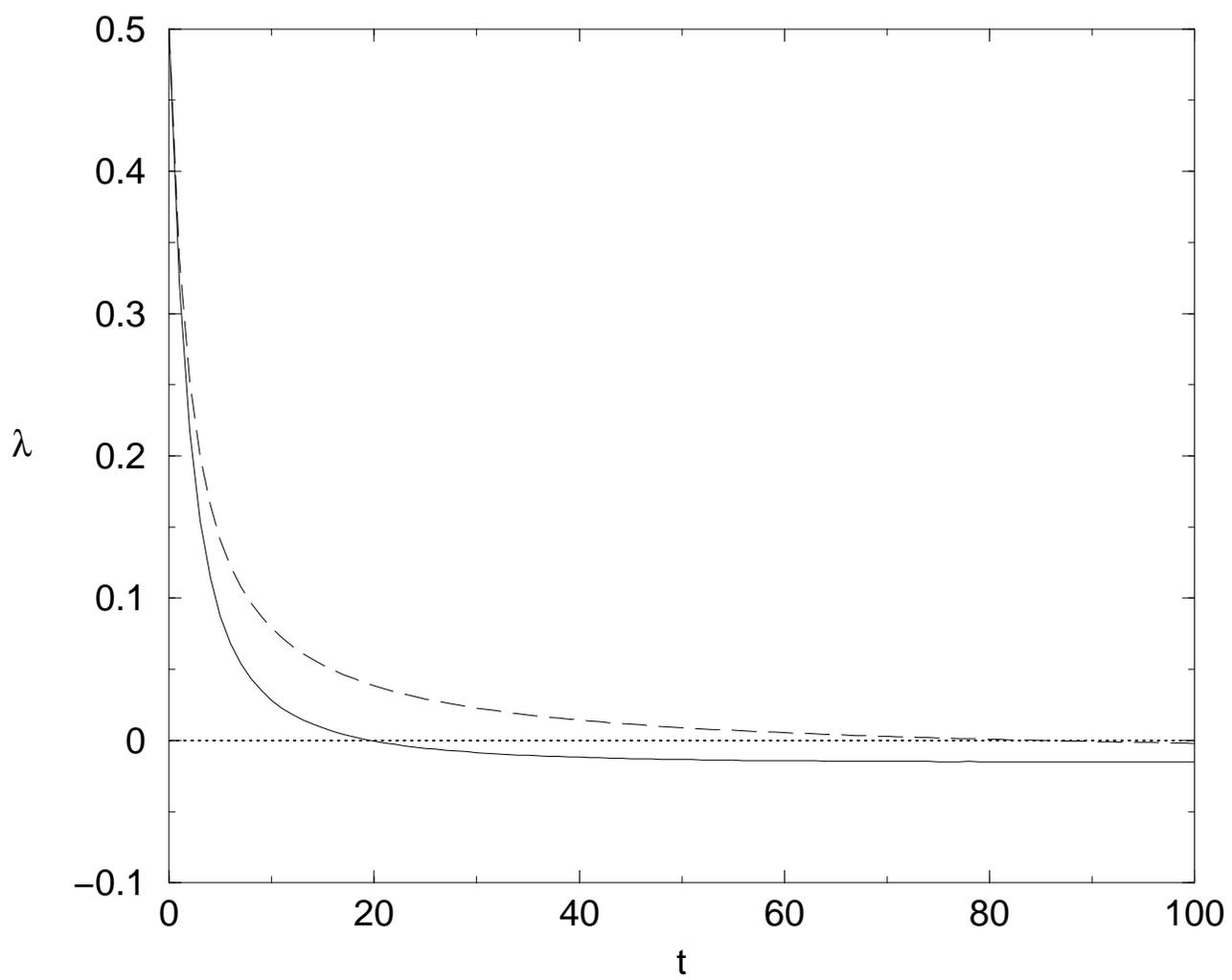
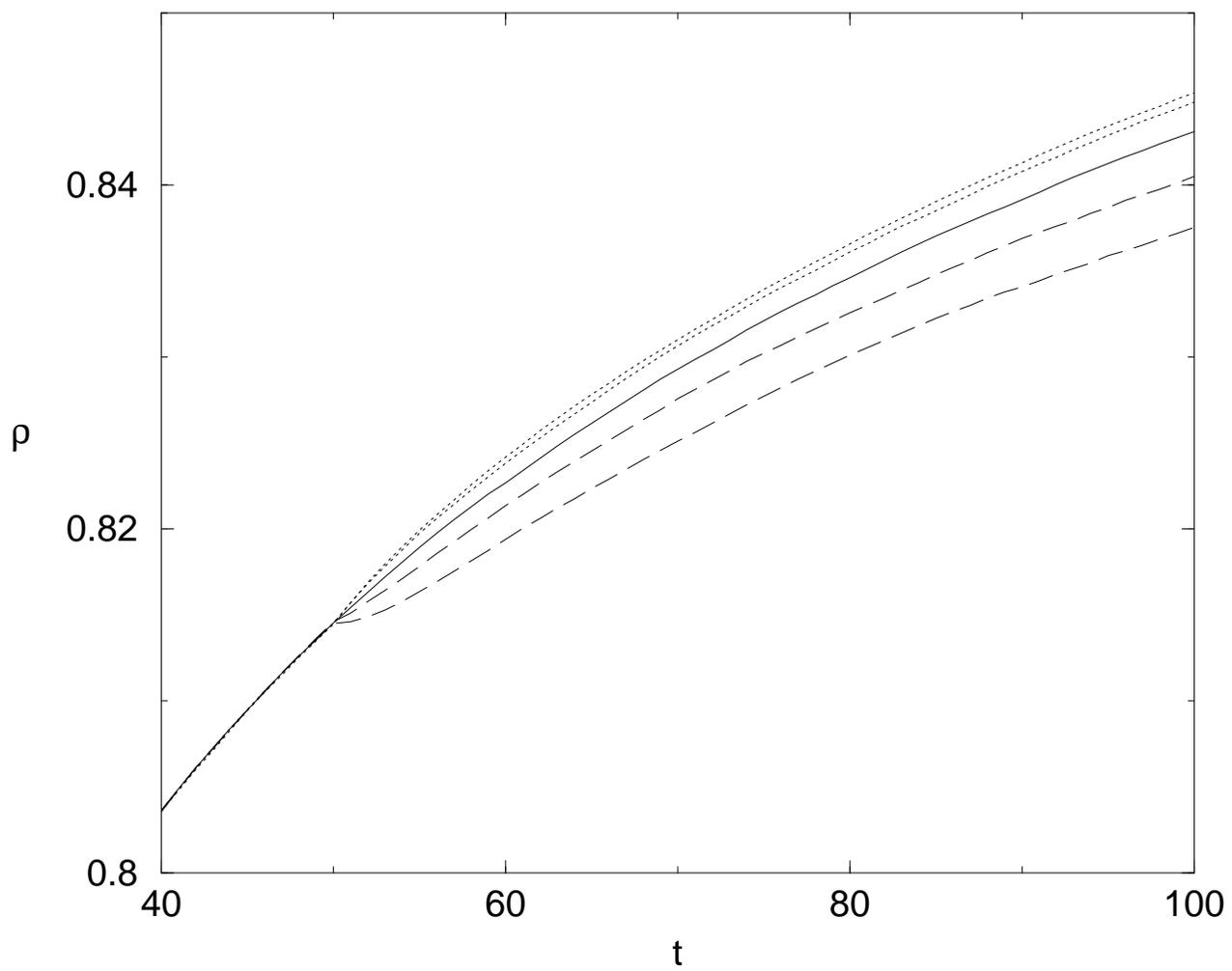
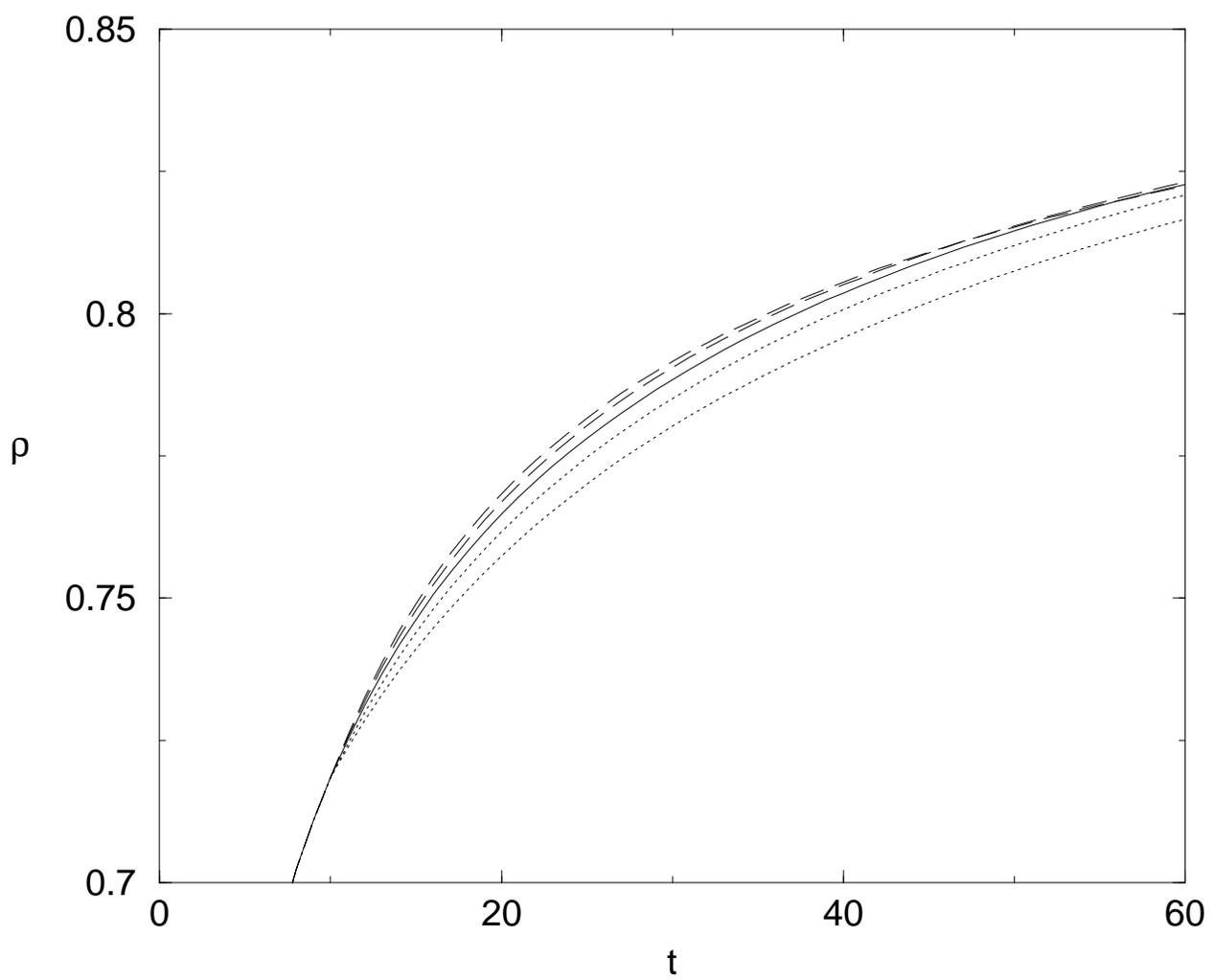
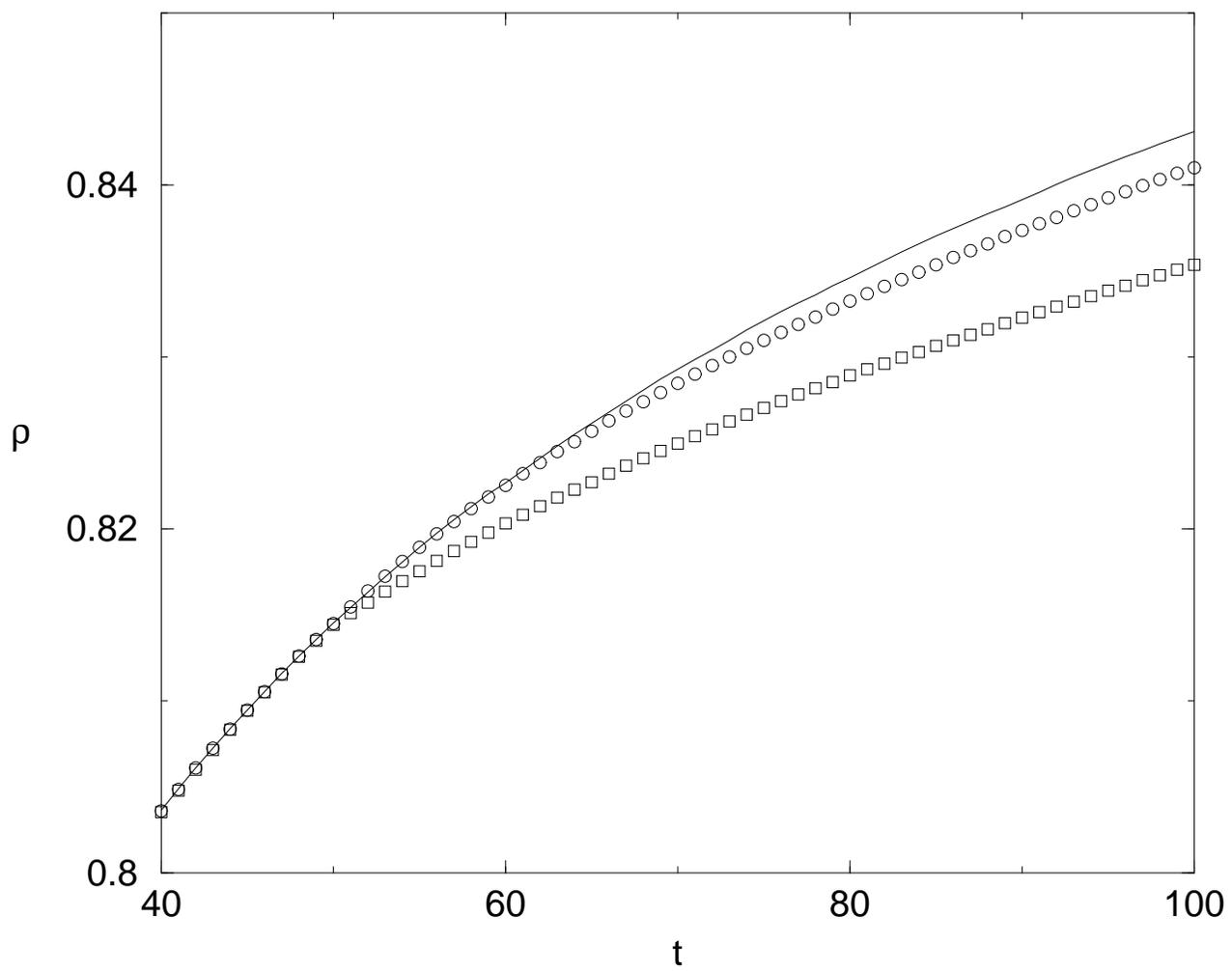
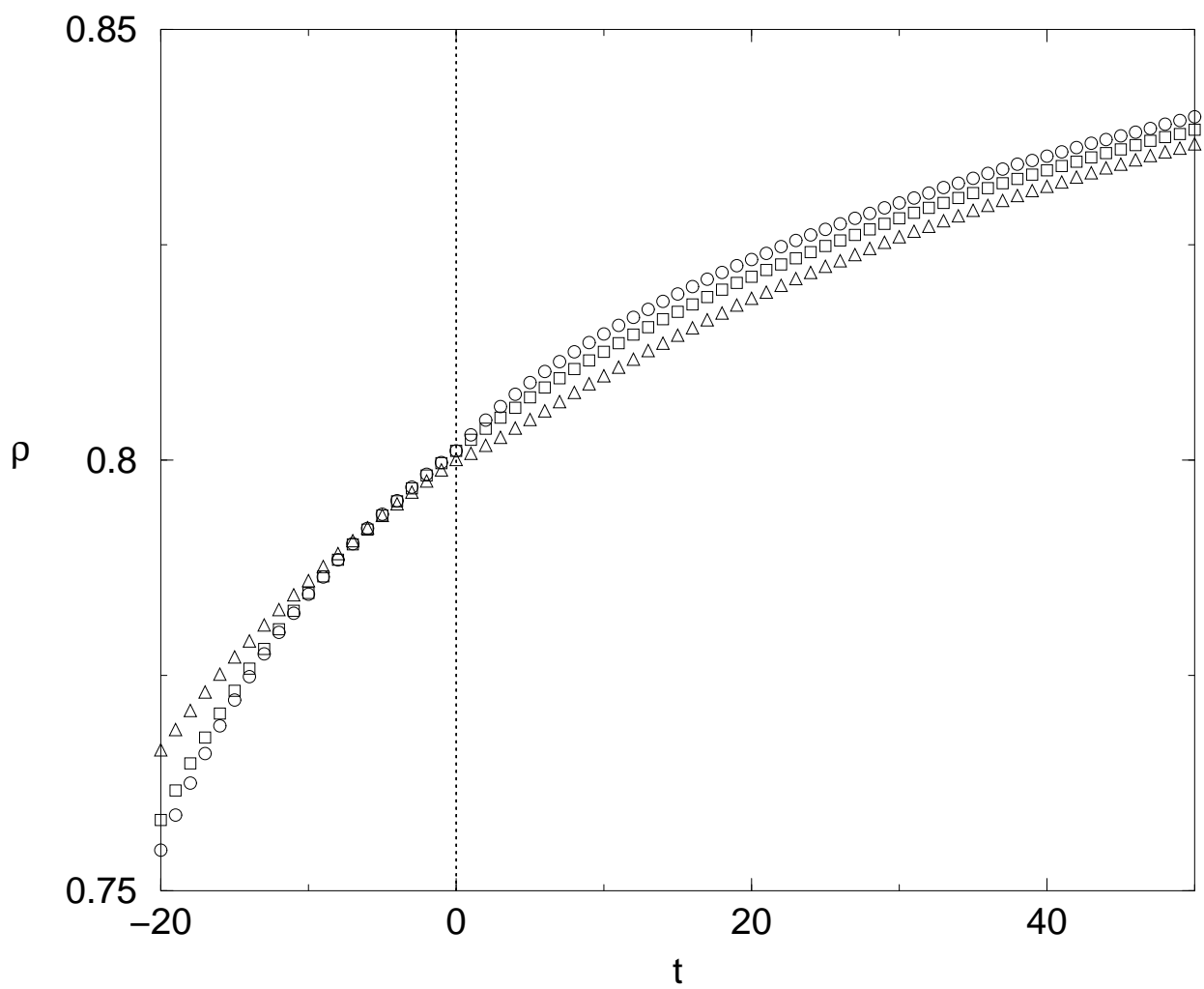
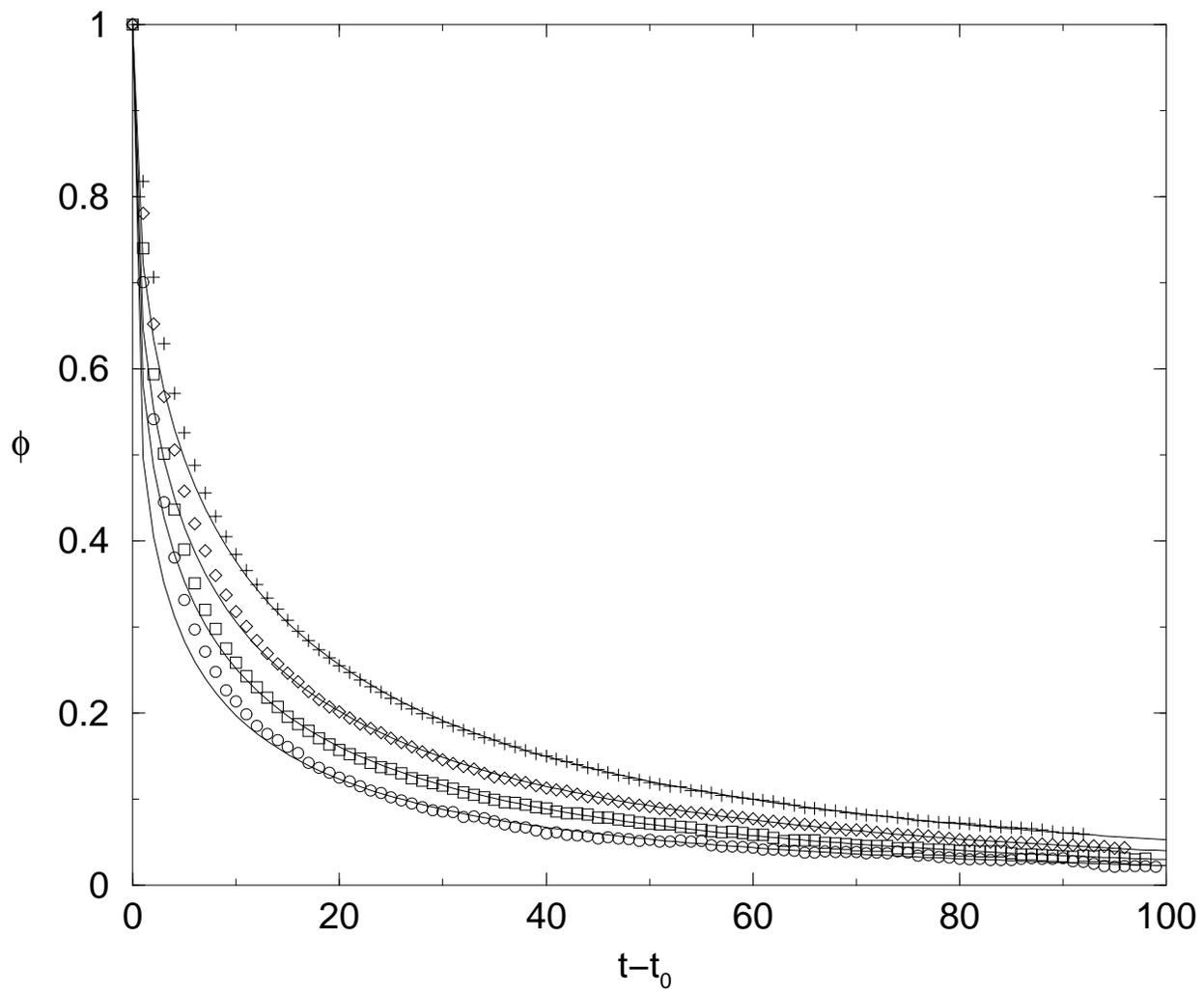
\end{document}